\documentclass{article}
\usepackage{times}
\usepackage{amsfonts}
\begin{document}
\noindent
{\Large ON THE VAN DER WAALS GAS, CONTACT GEOMETRY AND THE TODA CHAIN}
\vskip1cm
\noindent
{\bf D. Alarc\'on}$^{a}$, {\bf P. Fern\'andez de C\'ordoba}${}^{b}$, {\bf J.M. Isidro}${}^{c}$ and {\bf C. Orea}${}^{d}$\\
Instituto Universitario de Matem\'atica Pura y Aplicada,\\ Universidad Polit\'ecnica de Valencia, Valencia 46022, Spain\\
${}^{a}${\tt diealcor@doctor.upv.es}, ${}^{b}${\tt pfernandez@mat.upv.es},\\
${}^{c}${\tt joissan@mat.upv.es} , ${}^{d}${\tt carorhue@etsii.upv.es}\\
\vskip.5cm
\noindent
{\bf Abstract} A Toda--chain symmetry is shown to underlie the van der Waals gas and its close cousin, the ideal gas. Links to contact geometry are explored.

\section{Introduction}\label{einfuehrung}

The contact geometry of the classical van der Waals gas \cite{CALLEN} is described geometrically using a 5--dimensional contact manifold ${\cal M}$ \cite{ARNOLD} that can be endowed with the local coordinates $U$ (internal energy), $S$ (entropy), $V$ (volume), $T$ (temperature) and $p$ (pressure). This description corresponds to a choice of the fundamental equation, in the energy representation, in which $U$ depends on the two extensive variables $S$ and $V$. One defines the corresponding momenta $T=\partial U/\partial S$ and $-p=\partial U/\partial V$. Then the standard contact form on ${\cal M}$ reads \cite{BRAVETTI1, MRUGALA3}
\begin{equation}
\alpha={\rm d}U+T{\rm d}S-p{\rm d}V.
\label{cinque}
\end{equation}
One can introduce Poisson brackets on the 4--dimensional Poisson manifold ${\cal P}$  (a submanifold of ${\cal M}$) spanned by the coordinates $S$, $V$ and their conjugate variables $T$, $-p$, the nonvanishing brackets being 
\begin{equation}
\{S,T\}=1, \qquad \{V,-p\}=1.
\label{fundbra}
\end{equation}
Given now an equation of state 
\begin{equation}
f(p,T,\ldots)=0,
\label{aeternum}
\end{equation}
one can make the replacements $T=\partial U/\partial S$, $-p=\partial U/\partial V$ in order to obtain
\begin{equation}
f\left(-\frac{\partial U}{\partial V},\frac{\partial U}{\partial S},\ldots\right)=0.
\label{coelum}
\end{equation}
In ref. \cite{NOI} we have called Eq. (\ref{coelum}) {\it a partial differential equation of state}\/ ({\it PDE of state}\/ for short). It plays a role analogous to that played by the Hamilton--Jacobi equation in classical mechanics \cite{ARNOLD}. With respect to the latter, however, there is one fundamental difference. While in mechanics the Hamilton--Jacobi equation is just one equation (regardless of the number of degrees of freedom), in thermodynamics we have one PDE of state per degree of freedom, because the defining equation of each momentum qualifies as an equation of state.

\section{The PDE's of state of the van der Waals gas}\label{opertherm}

Let us consider one mole of particles of van der Waals gas ({\it i.e.}\/, Avogadro's number $N$ of particles). The fundamental equation in the energy representation $U=U(S,V)$ reads \cite{CALLEN}
\begin{equation}
U(S,V)=U_0\left(\frac{V_0}{V-b}\right)^{2/3}\exp\left(\frac{2S}{3Nk_B}\right)-\frac{a}{V},
\label{raewe}
\end{equation}
with $U_0,V_0$ certain fiducial values; setting $a=0$ and $b=0$ one recovers the ideal gas.
The variables $T$ and $-p$, conjugate to $S$ and $V$, are
\begin{equation}
T=\frac{\partial U}{\partial S}=U_0\left(\frac{V_0}{V-b}\right)^{2/3}\exp\left(\frac{2S}{3Nk_B}\right)\frac{2}{3Nk_B}
\label{emptp}
\end{equation}
and
\begin{equation}
p=-\frac{\partial U}{\partial V}=\frac{2}{3}U_0\exp\left(\frac{2S}{3Nk_B}\right)\frac{V_0^{2/3}}{(V-b)^{5/3}}
-\frac{a}{V^2}.
\label{niente}
\end{equation}
Eqs. (\ref{emptp}) and (\ref{niente}) lead to the van der Waals equation of state
\begin{equation}
\left(p+\frac{a}{V^2}\right)(V-b)=Nk_BT
\label{banderbals}
\end{equation}
and the equipartition theorem: 
\begin{equation}
U(T,V)=\frac{3}{2}Nk_BT-\frac{a}{V}.
\label{eierkratzer}
\end{equation}
The first PDE of state follows from Eq. (\ref{banderbals}),
\begin{equation}
\left(\frac{\partial U}{\partial V}-\frac{a}{V^2}\right)(V-b)+Nk_B\frac{\partial U}{\partial S}=0,
\label{primapde}
\end{equation}
while from Eq. (\ref{eierkratzer}) we obtain the second PDE of state:
\begin{equation}
U-\frac{3}{2}Nk_B\frac{\partial U}{\partial S}+\frac{a}{V}=0.
\label{secondapde}
\end{equation}
When $a=0$ and $b=0$, the system (\ref{primapde}) and (\ref{secondapde}) correctly reduces to the corresponding system of PDE's for the ideal gas, obtained in ref. \cite{NOI}. One readily verifies that integration of the system (\ref{primapde}), (\ref{secondapde}) leads back to the fundamental equation (\ref{raewe}) we started off with.

\section{Relation to the Toda chain}\label{dato}

A succession of changes of variables in configuration space ${\cal C}$ (the submanifold of ${\cal M}$ spanned by the extensive coordinates $S,V$) will relate the fundamental equation (\ref{raewe}) for the van der Waals gas to the potential energy of the Toda chain \cite{TODA}. We define the new variables $S', V'$
\begin{equation}
S':=S, \qquad  V':=V-b, 
\label{rimatreog}
\end{equation}
and $s,v$
\begin{equation}
s:=\frac{S'}{Nk_B}, \qquad v:=\ln \left(\frac{V'}{V_0}\right),
\label{kangez}
\end{equation}
in terms of which the fundamental equation (\ref{raewe}) reads
\begin{equation}
U(s,v)=U_0\exp\left[\frac{2(s-v)}{3}\right]-\frac{a}{V_0{\rm e}^v+b}.
\label{basco}
\end{equation}
The transformations (\ref{rimatreog}) and (\ref{kangez}) are both diffeomorphisms: they can be inverted, regardless of the values of the van der Waals parameters $a,b$. However the final change of variables
\begin{equation}
x:=s-v,\qquad U_0\exp\left(\frac{2y}{3}\right):= \frac{a}{V_0{\rm e}^v+b}
\label{bllkoz}
\end{equation}
becomes singular when $a=0$. For the moment we proceed under the assumption that $a\neq 0$, so (\ref{bllkoz}) is invertible. Then the fundamental equation (\ref{basco}) becomes
\begin{equation}
U(x,y)=U_0\left[\exp\left(\frac{2x}{3}\right)-\exp\left(\frac{2y}{3}\right)\right]=W(x)-W(y),
\label{uequisy}
\end{equation}
where we have defined the new function
\begin{equation}
W(z):=U_0\exp\left(\frac{2z}{3}\right).
\label{rematar}
\end{equation}
The function $W(z)$ coincides with the potential function of the Toda chain \cite{TODA}; we have already encountered it in ref. \cite{NOI} in the context of the ideal gas. Since the latter has $a=0$, which causes the change of variables (\ref{bllkoz}) to be singular, one must proceed differently in this case. Instead of Eq. (\ref{bllkoz}), a nonsingular change of variables to consider for the ideal gas is
\begin{equation}
x':=s-v,\qquad y':=s+v.
\label{becene}
\end{equation}
As already seen in ref. \cite{NOI}, this yields a fundamental equation depending on $x'$, but not on $y'$:
\begin{equation}
U_{\rm ideal}(x')=W(x').
\label{idelea}
\end{equation}
On the other hand from ref. \cite{TODA} we know that, in the limit of small wave amplitudes, the time average of the momentum variable in a thermal  ensemble of Toda chains is directly proportional to the product of Boltzmann's constant $k_B$ times the temperature $T$ (see Eq. (3.20) of ref. \cite{TODA}, the right--hand side of which is independent of the lattice site $n$). We conclude that, {\it in the limit of small amplitudes, a thermal ensemble of waves in the Toda chain behaves exactly as an ideal gas}\/. 

Returning now to the van der Waals gas in Eq. (\ref{uequisy}), the new canonical momenta read
\begin{equation}
p_x=\frac{\partial U}{\partial x}=\frac{2}{3}W(x), 
\qquad p_y=\frac{\partial U}{\partial y}=-\frac{2}{3}W(y).
\label{memen}
\end{equation}
While the momentum $p_x$ is the same as for the ideal gas, the negative sign in $p_y$ can be traced back to the reduction in energy, with respect to the ideal case, due to the van der Waals parameter $a$. The PDE's of state read, in the new variables $x,y$,
\begin{equation}
\frac{\partial U}{\partial x}-\frac{2U_0}{3}\exp\left(\frac{2x}{3}\right)=0, \qquad \frac{\partial U}{\partial y}+\frac{2U_0}{3}\exp\left(\frac{2y}{3}\right)=0.
\label{pdeseparat}
\end{equation}
Compared to Eqs. (\ref{primapde}) and (\ref{secondapde}) we see that, in the new variables $x,y$, the PDE's of state decouple into a system of two identical equations (up to a sign), one for each independent variable. Moreover, the equation corresponding to the variable $x$ equals that PDE of the ideal gas which expresses the equipartition theorem. Finally the contact form (\ref{cinque}) reads, in terms of $x,y$ and the corresponding momenta $p_x,p_y$,
\begin{equation}
\alpha={\rm d}U+p_x{\rm d}x+p_y{\rm d}y,
\label{contatto}
\end{equation}
which tells us that {\it the composition of all three diffeomorphisms (\ref{rimatreog}), (\ref{kangez}) and (\ref{bllkoz}) is a contactomorphism}\/ \cite{ARNOLD} (as had to be the case for a change of variables in configuration space ${\cal C}$). In the limit when the gas is ideal, the momentum $p_y$ vanishes identically \cite{NOI}, and the physics is described in terms of the 3--dimensional contact submanifold ${\cal N}$ spanned by $x,p_x$ and $U$.

\section{Discussion}

The physics of the classical van der Waals gas is usually described by a 5--dimensional contact manifold ${\cal M}$ endowed with the contact form given in Eq. (\ref{cinque}). In this paper we have identified one particular contactomorphism that neatly disentangles the (rather abstruse) fundamental equation (\ref{raewe}) to the much more manageable form given by Eqs. (\ref{uequisy}), (\ref{rematar}). This latter form is not just easier to work with; it is also more inspiring. Namely, the fundamental equation of the van der Waals gas now equals the difference of two terms (one term per independent variable $x,y$), each one of which is a copy of the Toda potential function \cite{TODA}. 

{}From the point of view of contact geometry, the only difference between the van der Waals gas and the ideal gas lies in the fact that the  contact manifold describing the van der Waals gas remains 5--dimensional, instead of reducing to the 3--dimensional contact submanifold ${\cal N}$ we found in the ideal case \cite{NOI}. Yet, as we have proved in Eq. (\ref{uequisy}), the fundamental equation can be expressed in terms of the Toda potential function in both cases. 

An intriguing feature of the above correspondence between the fundamental equation of a gas (either ideal or van der Waals) and the Toda potential function is the following. The small--amplitude limit considered in ref. \cite{TODA} is the limit of vanishing kinetic energy; this fact is reflected in the vanishing (to first order of approximation) of the time average of the generalised velocities $\dot s_n$ in ref. \cite{TODA}. This limit has been called {\it the topological limit}\/ in ref. \cite{NOS}; roughly speaking, it amounts to cancelling the kinetic term while keeping only the potential term in the Hamiltonian. This fact allows us to sharpen our previous correspondence, which we can now state more precisely as follows: {\it the classical thermodynamics of the (ideal or van der Waals) gas has a dual theory which, to first order of approximation, coincides with the topological limit of a thermal ensemble of waves in the Toda chain}\/.  Surprising here is the fact that, for the ideal gas, all energy is purely kinetic; and the potential energies introduced by the van der Waals parameters $a,b$ are almost negligible compared to the kinetic energy. Thus {\it the theory of gases, where energies are completely kinetic, or almost, is mapped by this correspondence into a dual theory in which kinetic energies are negligible}\/. Vanishing or at least negligible kinetic energies are strongly reminiscent of topological field theory \cite{SCHWARZ}; we hope to report on this issue in the future, as well as on its relation to Riemannian fluctuation theory \cite{RUPPEINER, LUISBERIS}.

\vskip.5cm
\noindent
{\bf Acknowledgements}  J.M.I. wishes to thank the organisers of the congress {\it Entropy 2018: From Physics to Information Sciences and Geometry}\/, Barcelona, for the opportunity to present a preliminary version of this work. This research was supported by grant no. ENE2015-71333-R (Spain) and Convocatoria Abierta 2015 para Cursar Estudios de Doctorado, SENESCYT (Ecuador).

\end{document}